\documentstyle[aps,prb,twocolumn,epsf]{revtex}
\begin{document}
\input{psfig.sty}
\draft
\newcommand{\lscoo}{\mbox{${\rm La_{1.5}Sr_{0.5}CoO_4}$}}
\newcommand{\cotwo}{Co$^{2+}$}
\newcommand{\cotri}{Co$^{3+}$}
\newcommand{\btau}{\mbox{\boldmath$\tau$}}
\newcommand{\bepsilon}{\mbox{\boldmath$\varepsilon$}}
\newcommand{\bQ}{\mbox{\boldmath$Q$}}
\newcommand{\bq}{\mbox{\boldmath$q$}}
\newcommand{\ba}{\mbox{\boldmath$a$}}
\newcommand{\bb}{\mbox{\boldmath$b$}}
\newcommand{\bc}{\mbox{\boldmath$c$}}
\newcommand{\br}{\mbox{\boldmath$r$}}
\newcommand{\h}{\mbox{$\frac{1}{2}$}}
\newcommand{\th}{\mbox{$\frac{3}{2}$}}

\twocolumn[\hsize\textwidth\columnwidth\hsize\csname
@twocolumnfalse\endcsname

\title{Spin-entropy driven melting of the charge order in La$_{1.5}$Sr$_{0.5}$CoO$_4$. }

\author{I.~A.~Zaliznyak$^{1}$, J.~M.~Tranquada$^{1}$,
 R.~Erwin$^{2}$, Y.~Moritomo$^3$}
\address{
 $^1$Department of Physics, Brookhaven National Laboratory, Upton,
 New York 11973-5000 \\
 $^2$National Institute of Standards and Technology, Gaithersburg,
 Maryland 20899\\
 $^3$Center for Integrated Research in Science and Engineering (CIRSE), Nagoya University,
Nagoya 464-8601, Japan
 }

\date{\today}
\maketitle

\begin{abstract}

We studied the melting of the charge order in the half-doped tetragonal
perovskite \lscoo\ by elastic neutron scattering. We found that diffuse
peaks, corresponding to the breathing-type modulation of oxygen
positions in the CoO$_6$ octahedra, disappear above $T_c = 825(27)$ K.
This melting of the diffuse superstructure is reversible (no change in
the correlation lengths upon annealing is observed) and accompanied by a
large nonlinear thermal expansion along the tetragonal $c$-axis, with a
cusp at $T_c$. We conclude that quenched disorder of the doped Sr ions
is at the origin of this {\it anisotropic charge glass} state. We also
suggest that its melting is driven by the transition from the
intermediate- to the high-spin state of Co$^{3+}$ ion, which is favored
by the spin entropy at high $T$.

\end{abstract}


]

Understanding the cooperative charge ordered phases in doped
transition-metal oxides is a key problem in the physics of
colossal-magnetoresistance (CMR) materials and high-temperature
superconductors.\cite{Orenstein2000,Tokura2000} Electron correlations in
these systems are driven by contributions from several equally important
interactions, such as Hund's coupling, splitting of the on-site
electronic levels by the crystal field, Jahn-Teller (JT) distortion,
Coulomb repulsion between the conduction electrons, superexchage, double
exchange, {\it etc}. Competing ground states lead to a rich variety of
phases, depending on the electronic configuration of the 3d metal ion,
its crystal environment, the doping level, and so on. To understand, and
ultimately predict, the nature and the physical properties of a given
transition-metal oxide at a particular doping, it is extremely important
to study the origin of, and the relation between, charge order (CO),
spin order (SO) and orbital order (OO) in different compounds.

The case of half-doping, which is relevant for the CMR materials, is of
special interest for several reasons. First, the spatial structure of
the CO instability is particularly simple -- it is a checkerboard-type
arrangement of the doped charges in the tetragonal plane. Second, it is
argued to be of most technological relevance in perovskite
manganites.\cite{Schiffer95} Finally, an orbital order, concomitant with
CO, was recently directly observed in several half-doped manganites by
X-ray resonant scattering.\cite{Murakami98,ZimmermanJirak} Consequently,
a number of theoretical papers appeared, aimed at explaining the origin
of the observed CO/OO at half-doping, its relation with spin order and
implications for the transport
properties.\cite{SolovyevBrink,vanDuin98,Natasha1999} To verify these
theories, and eventually achieve clear understanding of this important
problem, more experimental input is vital.

In a recent neutron scattering study \cite{Zaliznyak2000} we
characterized spin and charge order in the half-doped cobaltate \lscoo.
We found that the spins order at $T_s\approx 30$ K in a
quasi-two-dimensional incommensurate glassy structure, while
checkerboard short-range CO persists to at least $\sim 600$ K and shows
no measurable anomaly at $T_s$, {\it i.e.} the freezing of charge and
spin correlations are independent instabilities, inherent to the
electronic and crystal structure of \lscoo. Both magnetic and structural
scattering were similar to those observed in the isostructural manganite
La$_{0.5}$Sr$_{1.5}$MnO$_4$,\cite{Sternlieb96} where CO and SO
transition temperatures are much closer, and corresponding instabilities
are often presumed to be closely related. The main difference in our
case is the much shorter coherence range: correlations within the
tetragonal \ba-\bb\ planes extend only over $\xi_{\rm s}\sim 80$ \AA\
for SO, and less than $\xi_{\rm co}\sim 30$ \AA\ for CO, while $\xi_{\rm
co}\gtrsim 200$ \AA\ in
La$_{0.5}$Sr$_{1.5}$MnO$_4$.\cite{Wakabayashi2001} Despite its
short-range glassy nature, CO in \lscoo\ has no less drastic
consequences for transport properties than that in manganite,
\cite{Sternlieb96} resulting in an insulating state, and an activation
behavior of the electrical conductivity with $E_a\sim 6000$ K.
\cite{Moritomo98} In fact, such an apparently large activation energy
({\it i.e.} a weak temperature dependence of the conductivity), is
observed in the pseudo-cubic perovskite cobaltites RCoO$_3$,
R=La,Pr,Nd,Sm,Eu,Gd at temperatures $400-600$ K, \cite{Yamaguchi1996}
where a semiconductor-to-metal cross-over occurs upon heating.
Comparative studies of the isostructural Co and Mn compounds give a
unique opportunity to understand the relative importance of the
anisotropic crystal field energy, which in the cobaltites can be larger
than the on-site Hund's coupling, and the inter-site spin interaction,
which is often quenched for the Co$^{3+}$
ions.\cite{Zaliznyak2000,Asai98}

Here we focus our attention on the nature of the CO glassy state and the
charge melting process in \lscoo. We study the temperature evolution of
the CO correlation lengths and the charge order parameter up to and
above $T_c$, and investigate the effect of annealing and slow cooling of
the sample through the CO transition. Such procedure could distinguish
between the cooling-rate-dependent disorder of the mobile ions
(non-stoichiometric oxygen) or charge carriers, which can reorder into
an equilibrium arrangement,\cite{Wells1997} and that related to the
quenched disorder of the immobile Sr$^{2+}$ ions.

We studied the same high quality single crystal sample of \lscoo\ as in
Ref. \onlinecite{Zaliznyak2000}. At all temperatures the crystal
remained in the tetragonal ``HTT'' phase (space group $I4/mmm$), with
low $T$ lattice parameters $a=3.83$ \AA\ and $c=12.5$ \AA, Fig.~1(a). In
Ref. \onlinecite{Zaliznyak2000} we used the ``orthorhombic'' indexing,
based on the space group $F4/mmm$ with a unit cell that is two times
larger. This accounts for breaking of the translational symmetry in the
\ba-\bb\ plane by the checkerboard CO superstructure, and is convenient
for describing the spin order. Here, however, we will use the ``HTT''
$I4/mmm$ notation, \cite{Sternlieb96} which is relevant for the
charge-disordered phase, and is more appropriate for characterizing the
CO instability.

%==============================Fig.1==================================
\begin{figure} \noindent\vspace{-0.15in}
\parbox[b]{3.4in}{\psfig{file=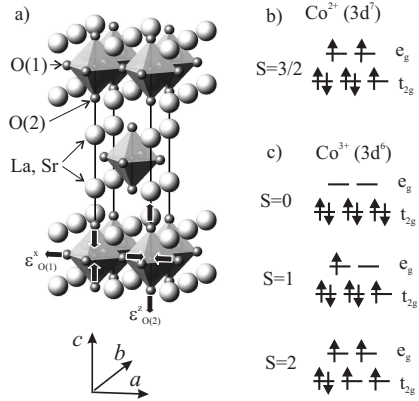,height=2.15in}
\vspace{0.2in} \noindent \caption{(a) Crystal structure of
La$_{1.5}$Sr$_{0.5}$CoO$_4$, arrows show the directions of oxygen
displacements \bepsilon$_{O(1),O(2)}$ accompanying the CO modulation.
(b),(c) Filling of the electronic levels in (b) Co$^{2+}$ with S=3/2,
and (c) in Co$^{3+}$ in the LS (S=0), IS(S=1) and HS (S=2) states.}}
\end{figure} \vspace{-0.1in} \noindent
%=====================================================================

Experiments were performed on BT2, BT4 and BT9 thermal beam 3-axis
neutron spectrometers at NIST Center for Neutron Research. PG(002)
reflections were used at the monochromator and analyzer, supplemented by
two PG filters to suppress the contamination from the higher order
reflections. The energy of the scattered neutrons was fixed at
$E_f=14.7$ meV, beam collimations were $80'-42'-62'-100'$ on BT2 and
BT4, and $40'-56'-51'-100'$ on BT9. The cylindrical sample was mounted
either in a displex refrigerator or in a 850~K vacuum furnace (BT9) with
the axis vertical, allowing wavevector transfers in the $(hhl)$
reciprocal lattice plane. Some scans on BT9 were also done at ambient in
the $(hk0)$ plane. Sample mosaic is $\lesssim 0.25^\circ$ in the $(hhl)$
plane, and $\lesssim 0.4^\circ$ in the $(hk0)$ plane. Normalization of
the scattering intensity was performed using the incoherent scattering
from a vanadium standard.

An extensive survey of elastic scattering from the super-structure
induced by the CO is presented in Fig. 2. It consists of broad
commensurate peaks centered at $Q = (h+0.5,k+0.5,l)$ reciprocal lattice
units (rlu) with $h,k,l$ integer, which corresponds to a checkerboard
arrangement of the \cotwo/\cotri\ valence in the \ba-\bb\ plane. An
increase in the intensity at large $Q$, characteristic of the scattering
from a structural modulation, is apparent.

%==============================Fig.2==================================
\begin{figure}[t] \noindent\vspace{-0.1in}
\parbox[b]{3.4in}{\psfig{file=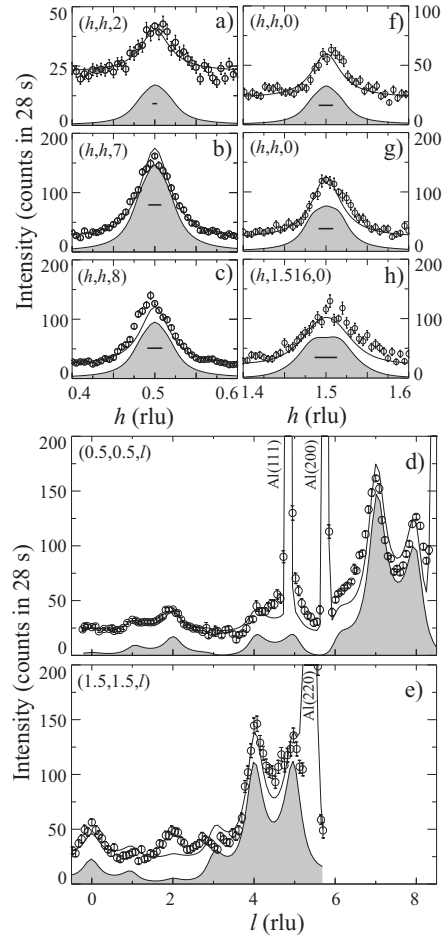,height=4.9in}
\vspace{0.15in} \noindent \caption{Elastic scattering from the diffuse
structural modulation accompanying charge order in \lscoo. (a)-(f) scans
collected on BT2 at $T = 6$ K in the $(hhl)$ reciprocal plane. (g),(h)
scans collected on BT9 at ambient $T\approx 295$ K in the $(hk0)$
reciprocal plane. Horizontal bars show the full width at half maximum
(FWHM) of the spectrometer elastic resolution along the scan direction.
Solid curves through the data are the resolution corrected fits to the
single peak cross-section (\ref{cross-section}). Shaded peaks show the
net intensity which would result from four incommensurate peaks at $Q_c
= \{ (0.5 \pm \epsilon,0.5 \pm \epsilon,l), (0.5 \pm \epsilon,0.5 \mp
\epsilon,l)\}$ with $\epsilon = 0.016$. }}
\end{figure} \vspace{-0.1in}
%=====================================================================

\noindent In all scans the widths of the peaks are substantially larger
than the instrument elastic resolution, which reveals the short-range
nature of the CO. The peaks are broader in the \bc-axis direction, so
that the super-structure is an anisotropic glass with the correlation
extending by about 4 lattice repeats in the \ba-\bb\ plane, and only
$2/3$ of a lattice repeat along the \bc\ axis, {\it i.e.} only between
the two nearest CoO$_4$ planes. The intensity modulation with $l$ shown
in the Fig. 2 (d), (e), exhibits a rather peculiar long-periodic
pattern, which translates into a modulation of a small object in real
space. In Ref. \onlinecite{Zaliznyak2000} we showed that scattering from
a breathing-type distortion of the oxygen octahedra surrounding the Co
ions provided a good account of a smaller set of data. This involves
displacements of the in-plane oxygens $\varepsilon^{x,y}_{O(1)}$, and
apical oxygens $\varepsilon^{z}_{O(2)}$, along the corresponding Co-O
bonds, as shown in Fig. 1 (a). The neutron scattering elastic
cross-section arising from such a modulation, short-range-periodic in
the crystal lattice, has the form of the factorized ``lattice
Lorentzians'', \cite{Fisher}

\begin{eqnarray}
\label{cross-section}
 \frac{d \sigma}{d \Omega} (\bq) = N I_{DW} \left| F(\bq) \right|^2
 \frac{1}{n_{\pm}}\sum_{+,-} {\cal L}_a^{\pm}(\bq)
 {\cal L}_b^{\pm}(\bq) {\cal L}_c^{\pm}(\bq)\;, \\
 {\cal L}_{a}^{\pm}(\bq) = \frac{\sinh{\xi^{-1}_a}}{\cosh{\xi^{-1}_a}
 - \cos{(\bq \pm \bQ)\cdot \ba}}\;, \;\;\;\; etc.
\end{eqnarray}
Here $F(\bq) = \sum_\mu (\bq\cdot\bepsilon_\mu) b_\mu e^{-i(\bq \cdot
\br_\mu)}$, $N$ is the number of unit cells in the crystal, $I_{DW}$ is
an effective Debye-Waller intensity prefactor, \bepsilon$_\mu$ is the
displacement of the atom $\mu$ inside a unit cell from its nominal
position $\br_\mu$, $b_\mu$ is its scattering length, \bQ\ is the
modulation wavevector, $\xi_a = \xi_b = \xi_{h}$ and $\xi_c = \xi_{l}$
are the in-plane and inter-plane correlation lengths measured in lattice
units (lu), and $n_{\pm}$ is the number of terms in the last summation.

We performed a global fit of all measured intensities at each $T$ with
Eq. (\ref{cross-section}), corrected for the instrument resolution, and
implying an equal-weight superposition of the commensurate modulations
with $\bQ = (0.5,0.5,l)$ for $l=0,1$. This corresponds to an equal
probability for the in-phase and anti-phase stacking of the checkerboard
CO structure for neighboring planes along \bc. At low temperatures the
Debye-Waller factor was fixed at $I_{DW}=1$, and the displacements of
the O nuclei from the nominal positions  $r_{O(1)} =
(0.5,0,0),(0,0.5,0)$ lu and $r_{O(2)} = (0,0,0.173)$ lu, the correlation
lengths $\xi_{h}$ and $\xi_{l}$, the background from the \bq-independent
incoherent elastic scattering and the Bragg scattering by the
polycrystalline aluminum in the sample environment were refined. Adding
a modulation of the La/Sr position, $r_{La,Sr} = (0.5,0.5,0.141)$ lu, as
well as varying $z_{O(2)}$ and $z_{La,Sr}$, did not improve the quality
of the fit. The solid lines through the data in the panels of Fig. 2
result from Eq. (\ref{cross-section}) with parameters obtained in the
global fit and summarized in the Table 1. Clearly, the scattering from
the oxygen displacements gives a very good description of all measured
intensities.

In the Ref. \onlinecite{Zaliznyak2000} we also found magnetic elastic
scattering which appears below $T \approx 30$ K at slightly
incommensurate positions $Q_m = (\frac{0.5 + \epsilon}{2}, \frac{0.5 +
\epsilon}{2}, l)$ with $\epsilon \approx 0.016$ and $l$ odd-integer. If
this incommensurability of the spin order is the result of a stripe
order of the doped charges, related to a small deviation of the La:Sr
ratio from the nominal 1.5:0.5, or some non-stoichiometric oxygen, we
would also expect four incommensurate CO peaks at $Q_c = \{ (0.5 \pm
\epsilon,0.5 \pm \epsilon,l), (0.5 \pm \epsilon,0.5 \mp \epsilon,l)\}$,
as observed in the doped cuprates. \cite{Tranquada} However, for the
scans in the $(h,h,l)$ plane, Fig. 2 (a)-(f), the major contribution of
the out-of-plane peaks, accepted by the coarse vertical resolution of
the spectrometer and centered at $h=0.5$, would hinder the experimental
observation of the incommensurate splitting. \cite{Zaliznyak2000} To
clarify this issue we made a number of scans through the CO scattering
around $(1.5,1.5,0)$ and $(2.5,0.5,0)$ in the $(h,k,0)$ plane. No
splitting was apparent in any of these scans either, as illustrated by
two representative datasets in Fig. 2 (g),(h). The overall peak width is
always dominated by the short correlation length, which smears the
implied small splitting. In other words, the correlation range of the
CO, $\xi_{h} \approx 4$ lu, is so small compared to the wavelength of
the expected incommensurate modulation $\lambda_\epsilon = 1/\epsilon
\approx 60$ lu, that the latter is undetectable. The cross-section
resulting from four incommensurate peaks with $\epsilon = 0.016$ is
shown by the shaded areas in the Fig. 2. It gives an almost identical
fit to the measured intensities, with only a slightly larger $\xi_{h}$
to ``compensate'' for the splitting, see Table 1.

%==============================Fig.3==================================
\begin{figure}[t] \noindent\vspace{-0.4in}
\parbox[b]{3.4in}{\psfig{file=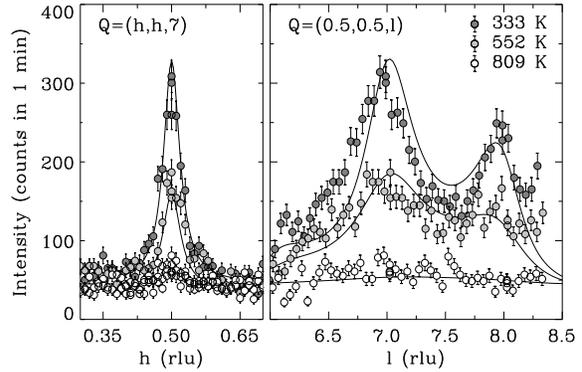,height=2.75in}
\vspace{-0.5in} \noindent \caption{Change in the diffuse
super-structural scattering concomitant with melting of the CO in
\lscoo. }}
\end{figure} \vspace{-0.1in}
%=====================================================================

Figure 3 shows the evolution of the diffuse CO scattering upon heating.
Although a small peak may still be present in the $h$-scan at $T=809$ K,
we find that it completely disappears at 850 K. Each measurement at high
$T$ was preceded by a sample re-alignment, since the melting of the
charge order in \lscoo\ is accompanied by an anomalously large thermal
expansion along the \bc-axis, Fig. 4(a). We find that the temperature
dependence $c(T)$ saturates somewhere above $800$ K, with a kink
indicative of a second order phase transition. This supports our
suggestion \cite{Zaliznyak2000} that the decrease in the $c$ lattice
spacing below $T_c$ is proportional to the charge order parameter
$\eta_{c}$. Fitting it with the expression $c(T) = c(850\,K) - \alpha
(T_c - T)^\beta$ we get $T_c = 825(27)$ K and $\beta = 0.59(15)$,
indistinguishable from the mean-field value 0.5. The in-plane lattice
constant $a$, on the other hand, stays almost unchanged, with some
tendency to decrease upon heating, and exhibits no anomaly at $T_c$.
These findings clearly indicate that electrons involved in the CO are
localized in the out-of-plane $d_{3z^2-r^2}$ orbitals.

Interestingly, to within our errors the width of the peak in $h$ does
not change upon heating, i.e. the in-plane correlation length stays
constant, and CO melting mostly involves the loss of the inter-plane
coherence with increasing $T$, Fig.4 (b). In the fits of the scans for
$T>300$ K we fixed the oxygen displacements and varied the ``Debye-
Waller'' prefactor to account for the decrease in the CO scattering
intensity, which follows the disappearance of

%==============================Fig.4==================================
\begin{figure}[t] \noindent\vspace{-0.1in}
\parbox[b]{3.4in}{\psfig{file=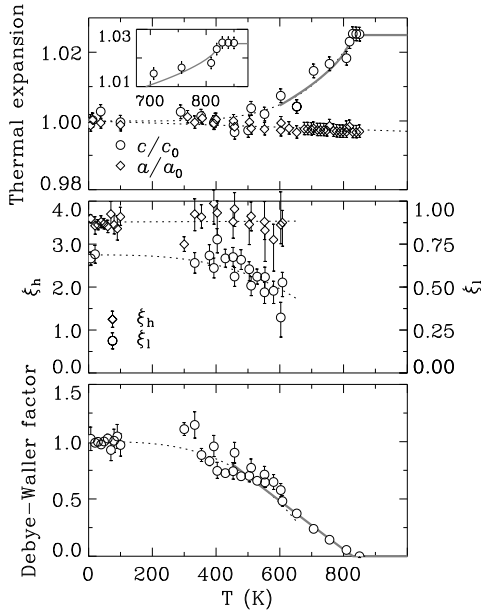,height=3in}
\vspace{0.4in} \noindent \caption{Temperature dependence of the
structural parameters related to the CO in \lscoo. (a) change of the
lattice constants relative to the low-$T$ values. (b) in-plane (left
scale) and inter-plane (right scale) correlation length, for $T\geq 650$
K they were fixed at $\xi_h = 3.5$ lu, $\xi_l = 0.32$ lu. (c) intensity
(``Debye-Waller'') prefactor. Broken curves are the guides for the eye,
solid lines are the critical-law fits (see text). }}
\end{figure} \vspace{-0.1in}
%=====================================================================

\noindent the order parameter $\eta_{c}$, Fig. 4(c). The solid line in
Fig. 4(c) is the best fit to the expression $I_{DW}(T) \sim (T_c -
T)^\gamma$, which gives $T_c = 821(48)$ K, identical with that refined
from the $c(T)$, and $\gamma = 0.92(11)$. The exponent $\gamma$ agrees
well with $\gamma = 2\beta$, expected from $I_{DW} \sim \eta_{c}^2$, and
also with the mean-field value $\gamma = 1$.

We found that upon slow cooling, even after annealing the sample for a
few hours at $T=850$ K, the CO peaks reappear below $T_c$ with the same
width and intensity as before. The reversible nature of the melting
transition is important for understanding the origin of the {\it charge
glass} phase in \lscoo.  It suggests that the quenched disorder of the
dopant Sr ions, which strongly interact with the doped charges, is the
most probable cause for the loss of the CO in-plane coherence, and not a
non-stoichiometric oxygen or rapid quench through $T_c$.

A key for understanding the \cotri/\cotwo\ mixed valence phases and
cobaltites in general is provided by the peculiar structure of the $3d$
electron levels. In a cubic crystal field $3d$ levels are split into a
lower $t_{2g}$ triplet and an $e_{g}$ doublet of $d_{x^2-y^2}$ and
$d_{3z^2-r^2}$ orbitals. In cobalt the Hund splitting of spin ``up'' and
``down'' $3d$ levels is anomalously small, and though Co$^{2+}$ ($3d^7$)
is in the ``normal'' S=3/2 state $t_{2g}^5e_g^2$, Fig. 1(b), Co$^{3+}$
($3d^6$) may be found in either the high $t_{2g}^4e_g^2$ (HS, S=2),
intermediate $t_{2g}^5e_g^1$ (IS, S=1), or low spin state $t_{2g}^6$
(LS, S=0), Fig.~1(c). If LS were the ground state, as in LaCoO$_3$, a
decrease in the free energy $F=E-T\ln({\rm 2S+1}$) due to the higher
paramagnetic entropy may drive transitions to IS and HS states with
increasing temperature. \cite{Asai98,Saitoh1997} On the other hand, it
is clear from Figure 1(c) that only the IS S=1 state is JT-active, {\it
i.e.} favors tetragonal splitting of the $e_{g}$ doublet, as it has one
electron there. In either the HS or LS state, such splitting gains no
energy because both $e_{g}$ levels are equally occupied.

We conclude, that \cotri\ ions in \lscoo\ are in the IS state at low
$T$, which favors the JT-distorted CO phase. At higher $T$, a
spin-entropy driven transition to the HS state occurs (similar IS-HS
transition is observed in LaCoO$_3$ at $\approx 600$ K \cite{Asai98})
with consequent disappearance of the JT modulation and conspicuous
melting of the charge order. Therefore, it appears that local effects
such as the spin-state transition and JT level splitting are very
important and quite efficient in driving the CO phase in \lscoo, and,
perhaps, in other materials.

We thank NIST Center for neutron Research for hospitality, and
J.~P.~Hill for valuable remarks. This work was carried out under
Contract NO. DE-\-AC02-\-98CH10886, Division of materials Sciences, US
Department of Energy.

\begin{table}
\label{table1} \caption{Correlation lengths and oxygen displacements
(lu) obtained from the global fit of the measured CO scattering at
$T<300$ K with the Eq. (\ref{cross-section}), $\epsilon$ is an implied
incommensurability of the CO in the \ba-\bb\ plane (lu).}
\begin{tabular}{ccccc}
 $\epsilon$ & $\varepsilon^{x,y}_{O(1)}$ & $\varepsilon^{z}_{O(2)}$
 & $\xi_{h}$ & $\xi_{l}$ \\
\tableline
 0 & 0.012(1) & -0.0058(4) & 3.7(3) & 0.67(5) \\
 0.016 & 0.012(1) & -0.0057(3) & 4.9(6) & 0.65(4) \\
\end{tabular}
\end{table}

\end{document}